%% file: CommF1_5.tex
\documentclass[11pt]{article}
\usepackage{graphicx}
\usepackage[margin=1.25in]{geometry}
\usepackage[usenames,dvipsnames]{color}
\usepackage{url}
\usepackage[colorlinks = true,
            linkcolor = blue,
            urlcolor  = blue,
            citecolor = blue,
            anchorcolor = blue]{hyperref}

\newcommand\pubnumber{Transcendental Preprint }
\newcommand\pubdate{\today}

\def\Title#1{\begin{center} {\LARGE #1 } \end{center}}
\def\Author#1{\begin{center}{ \sc #1} \end{center}}

\newcommand\pubblock{\rightline{\begin{tabular}{l} \pubnumber\\
         \pubdate \end{tabular}}}
\newenvironment{Abstract}{\begin{quotation} \begin{center}
                       ABSTRACT
     \end{center}\bigskip  }{\end{quotation}}

\input workshopsymbols.tex

\newcommand\snowmass{\begin{center}\rule[-0.2in]{\hsize}{0.01in}\\\rule{\hsize}{0.01in}\\
\vskip 0.1in Submitted to the  Proceedings of the US Community Study\\ 
on the Future of Particle Physics (Snowmass 2021)\\ 
\rule{\hsize}{0.01in}\\\rule[+0.2in]{\hsize}{0.01in} \end{center}}

\begin{document}

\pubblock

\Title{Programs Enabling Deep Technology Transfer from National Labs}

\bigskip 

% \Author{$^1$}
\Author{\textbf{Yasou Arai$^1$, Farah Fahim$^2$, Ryosuke Furubayashi$^3$, Matthew Garrett$^5$, Shaorui Li$^2$, Kathleen McDonald$^6$, Aaron Sauers$^2$, Mauricio Suarez$^2\ast$, Koji Yoshimura$^4$\\}
$^1$High Energy Accelerator Research Organization (KEK), Japan \\
$^2$Fermi National Accelerator Laboratory, US \\
$^3$Institute of Physical and Chemical Research (Riken), Japan\\
$^4$Okayama University, Japan\\
$^5$SLAC National Accelerator Laboratory, US\\
$^6$CloudArc, US\\
$^\ast$E-mail: suarez@fnal.gov } 

\medskip

%\Address{Fermi National Accelerator Laboratory, Batavia, IL USA}

\medskip

\begin{Abstract}
To maximize the technology transfer potential, it is important to create an ecosystem for the inventors to adapt the technologies developed for basic science to successful commercial ventures. In this white paper we present a brief overview of technology transfer programs at high energy physics (HEP) laboratories with a focus on the programs at Fermilab and KEK, and identify opportunities and recommendations for increasing partnerships and commercialization at HEP-centric laboratories.
\end{Abstract}

\snowmass
\def\thefootnote{\fnsymbol{footnote}}
\setcounter{footnote}{0}

\pagebreak
\tableofcontents

\pagebreak

\section*{Executive Summary}
To maximize the technology transfer potential, it is important to create an ecosystem for the inventors to adapt the technologies developed for basic science to successful commercial ventures. In this white paper we present a brief overview of technology transfer programs at high energy physics (HEP) laboratories with a focus on the programs at Fermilab and KEK. Upon evaluating the opportunities,  seven major recommendations are present for increasing partnerships and commercialization at HEP-centric laboratories:
\begin{itemize}
     \item Aligning inventor royalty distribution policy across DOE
     \item Engaging with partnership intermediaries to accelerate commercialization
     \item Early identification of dual-use innovations
     \item Increasing technology transfer educational opportunities targeted to HEP researchers
     \item Public-private partnerships for accelerating HEP innovations
     \item Extending other transaction authority
     \item Establishing an entrepreneurial leave program
\end{itemize}

\section{Motivation}
DOE National labs are engaged in cutting edge research for next generation technologies to enable lofty goals set by experiments for basic science discovery. The next generation of DOE facilities for Colliders, Neutrinos, Astrophysics, etc. not only require demonstration of new concepts but also necessitate at-scale prototyping. While on one hand universities are focused on surpassing the state-of-art, industry on the other hand is focused on quality and repeatability. 

DOE experiments provide the motivation to go beyond proof of concept and manufacture at mid-scale, which can eventually lead to establishing a path for commercial production. Manufacturing for experiments increases the technology readiness level as well as making the process robust and cost effective. Hence DOE experiments and applications are a key driver of lab to fab innovation. 

In order to maximize the technology transfer potential, it is important to create an ecosystem where the technology inventors can create and sustain spin-offs/startups. Adapting the technologies developed for basic science to successful commercial ventures is a long and arduous process. In this white paper we evaluate the opportunities that can be made available to foster and sustain technology transfer.

\section{Example Commercialization Processes of HEP Technologies by Spin-offs or Start-ups}

\subsection{Fermilab Commercialization Process}
IP created by a Fermilab employee as part of their employment contract is owned by Fermi Research Alliance (FRA), the management and operating contractor at Fermilab. The inventor/employee benefits by sharing in the royalty proceeds from licensing the invention: the inventors get a percentage (after patent expenses) of the revenue received by FRA. 

For a national lab inventor to request a license from a national lab, after there has been an invention disclosure and the IP has been secured (filed patent application or asserted copyright) by the Office of Partnerships and Technology Transfer (OPTT) at Fermilab, the first step is for the inventors to connect with OPTT to discuss commercialization options. Some inventions are better suited for established companies. Other inventions are well suited for a startup. Startups usually reduce risk and uncertainty on a commercial space that is unclear at the time of the license.

A license requester should be aware of the difference between an “option to negotiate a license” and a “license”. An option to negotiate a license (or “Option") is a promise not to license to another person/entity during a specific amount of time (usually six months to a year). It is common for an inventor to first secure an Option with a specific goal (write a business plan, secure team, etc.) and then obtain a license to the technology.

Various licensing terms may be negotiated, including both exclusive and nonexclusive license grants. Exclusive licenses may be in certain fields of use, geographic areas, or according to other terms. Co-exclusive and partially exclusive licenses, where exclusive rights to commercialize a technology may be shared by several organizations or restricted by area of use, territory, or other terms, also may be granted. For example, one company may obtain exclusive rights to use an invention in the energy industry, while another exclusively licenses the same invention for application in the food industry.

FRA must perform an encumbrance review to ensure the IP is available for licensing generally and the requested license specifically. The Laboratory performs due diligence to ensure the proposed license arrangement does not impinge on its contractual obligations, such as existing license agreements or options to license.

More details of the Fermilab commercialization process can be found in Appendix A. 

\subsection{KEK Commercialization Process}
KEK is established as an Inter-University Research Institute, so there are many programs that enable university researchers' access to KEK technologies and expertise. Funding of major projects are coming directly from Ministry of Education, Culture, Sports, Science and Technology (MEXT) to KEK. KEK can afford travel expenses of university researcher, and the researchers can access most of KEK equipment, facility in principle.
KEK (Tsukuba campus) is located in Tsukuba Science City where 150 national and private laboratories/universities are accumulated. Thus, many kinds of collaborating activities exist. 

One of such activity is Tsukuba Innovation Arena (TIA, https://www.tia-nano.jp/page/\\dir000002.html). TIA is an open innovation hub operated by 6 public organizations: the National Institute of Advanced Industrial Science and Technology (AIST), the National Institute for Materials Science (NIMS), the University of Tsukuba, the University of Tokyo, Tohoku University and KEK. To drive innovation in Japan, these six stellar organizations of TIA collaborate and compile their resources for R\&D (e.g., researchers, facilities, and intellectual property) and support the creation of knowledge and its application in industry. TIA also fosters next-generation scientists and engineers. TIA is supported by Keidanren (Japan Business Federation).

Most technologies developed at KEK are published and accessible to the public, according to tradition in the high-energy physics fields. However, in recent days, protection of the technology with patents is required to promote collaboration with industry.

At first, the inventor has to report any inventions to the Intellectual Property and Partnership Office ``IPPO" at KEK. Then the invention is evaluated by an Intellectual Property committee. If it is determined that the invention is part of KEK’s employee’s contract and has commercial potential, then the right of the IP is transferred from the inventor to KEK. 

A non-exclusive license may be obtained by companies for KEK owned patents. If the patents are obtained in collaboration with a company and shared, exclusive license is permitted to the company for a limited period.

For the joint patent case, KEK requests non-implementation compensation and/or payment of cost related to the patent application and maintenance to the company.
It is very hard to get large amount of license fees. Instead of getting the license fees, we think KEK owned patents are beneficial to get external funding and promote collaboration with companies.
Since 2013, the government has placed University Research Administrators (URA) at KEK and many universities. URA attempts to connect inventions by researchers to industry sectors. 

More details of the KEK commercialization process can be found in Appendix B.
%\subsection{Collaboration with Large National/Regional Research Institutes and National Laboratories}

\section{Recommendations for increasing partnerships and commercialization}
%Fermilab has been considering modifying its inventor royalty sharing policy for many years, so a recommendation to do so here might make sense. The proposed modification would also more closely align with KEK's policy as well as the UC-managed DOE labs. Policy under consideration is 34\% to inventors, 33\% to the inventor's division, and 33\% to the laboratory for programs like technology maturation.
\subsection{Aligning Inventor Royalty Distribution Policy Across DOE}
Inventor royalty distribution policy varies across the DOE complex. Lawrence Livermore National Laboratory (LLNL) and Los Alamos National Laboratory (LANL) distribute 35\% of their net royalties to the inventors. Oak Ridge National Laboratory (ORNL) and Pacific Northwest National Laboratory (PNNL) distribute roughly 15\%, differing with respect to that which is netted out prior to dispensation. Sandia National Laboratories (SNL) distributes 20\% of its royalties to the inventor(s)~\cite{osti_889004}.
%I got LLNL, LANL, ORNL, and PNNL data from "Federal Technology Transfer Requirements:..." paper. Placed reference in references section.

Fermilab's policy directs 50\% of annual net royalties below \$10,000, and 25\% of net royalties in excess of \$10,000, to the inventor(s). This policy is favorable to the inventor(s) when expected annual royalties are modest. If, however, forecasted annual royalties are greater than \$10,000, the expected inventor share in licensing proceeds will approach 25\%. 

A proposed consistent royalty distribution might entail 34\% to the inventor(s), 33\% to the inventor's division, and 33\% to the Laboratory. The royalty portion distributed to the Laboratory or Division might be used to support HEP opportunities, consortium investments, rewards programs, technology maturation to further the technology transfer mission, and innovation/entrepreneurship educational opportunities for Laboratory staff.

%The following is from Matt Garrett
\subsection{Engaging with Partnership Intermediaries to Accelerate Commercialization}
The majority of US DOE national laboratories that support high-energy physics facilities are small multi-program or single purpose laboratories that have limited resources available to support Technology Transfer (TT) and commercialization activities. The resource allocations for TT staff are generally proportional to other administrative support functions at these laboratories. However, commercial impact and viability is unrelated to laboratory size. Disruptive technologies can be created regardless of lab funding, program portfolio diversity or technology transfer support staffing. 

A Partnership Intermediary (PI) is a non-profit entity with specialized skills that can assist federal agencies and laboratories in TT and commercialization functions. The use of a PI is authorized under US Title 15 U.S.C. §3715 and is defined as “a State or local government, or a nonprofit entity owned in whole or in part by, chartered by, funded in whole or in part by, or operated in whole or in part by or on behalf of a State or local government, that assists, counsels, advises, evaluates, or otherwise cooperates with small business firms, institutions of higher education” and “educational institutions.” 

Pilot programs funded in the past by the Department of Energy Office of Technology Transitions (OTT) ~\cite{uidp_2020} to evaluate how a PI could interface with high-energy physics funded national laboratories showed some promise in assessing technologies for market pull, marketing technologies which are innovated in high-energy physics research areas, and matchmaking technologies at national labs with entrepreneurs in private industry who are interested in taking high-energy physics innovations from the laboratory to the market.  These intermediaries can provide support to accelerate innovation and commercialization from laboratories.

\subsection{Early Identification of Dual-Use Innovations} 
Many of the innovations generated in high-energy physics facilities are developed and designed primarily to support facility construction and instrument design.  However, a number of these innovations have what are called “dual-uses”, meaning that the innovations have applications in several other fields and concepts of operation outside their intended use.  For example, physics-based AI algorithms developed to identify anomalies in large data streams from high-energy physics experiments may have a range of applications in data analytics and autonomy.  

Working with technology transfer entities, scientists and engineers developing new technologies can identify “dual-use” application cases where inventions identified early can proactively develop market analysis and case studies to support leveraging and commercializing technologies in areas beyond their immediate use.   Many of the barriers to capturing innovation from high-energy physics technology areas come from limited funds to pursue patents effectively, which requires technology transfer and laboratory leadership to evaluate innovations on their return on investment (ROI) and pull from the marketplace.    Developing value propositions and market analysis upfront on innovations with identified potential applications beyond their original intended use can accelerate acceptance and impact in the marketplace.

\subsection{Increasing Technology Transfer Educational Opportunities Targeted to HEP Researchers} 

A gap currently exists in many of the national laboratories in education of scientists and engineers working in high-energy physics facilities on the basics of technology transfer and intellectual property.  While national programs such as I-Corps are reaching many of the innovators in applied research areas where the link between innovation and market products are quite clear, more support is required for high-energy physics researchers to help them understand the relevance of their innovations in the broader marketplace.

Providing educational opportunities to ramp up to an I-Corps level of engagement for high-energy physics researchers would be a great opportunity to provide the building blocks to researchers to enable more engagement in capturing innovations.  An intermediate step may be adoption of an I-Corps "Lite" program which has demonstrated success at other labs where time commitments and fewer resources constrain participation in a full I-Corps program.  Where researchers and TT staff have sufficient interest and engagement to justify coverage of abbreviated I-Corps content, a "Lite" program may be a useful to step encourage thinking about intellectual property and technology commercialization concepts.  Discussions on the types of intellectual property (ex. patents, copyrights), rights afforded to researchers from their innovations, and the mechanisms to engage with industry to advance their technologies would provide valuable resources and perspectives

\subsection{Public-Private Partnerships for Accelerating HEP Innovations}

A major barrier with market proliferation of high-energy physics innovation is the decreasing pool of companies who have the business models, infrastructure, and capabilities to develop and manufacture the technologies evolving out of the laboratories. Specifically, in technology areas such as accelerators, US federal program managers  have proposed developing public-private partnerships to foster and support small and large technology businesses who collaborate with the laboratories and serve as commercialization partners for critical technologies developed as part of facilities and experiments in high-energy physics.   These public-private partnerships could serve as both advocacy and economic development entities for high-energy physics derived technologies, as well as matchmakers which aid companies and laboratories in forming collaborations which lead to commercialization outcomes.

 %from early OTA draft
\subsection{Extending Other Transaction Authority}
DOE has created a number of tools the labs can use to collaborate with industry, including Strategic Partnership Projects (SPPs), Cooperative Research and Development Agreements (CRADAs), User Facility Agreements, Agreements for Commercializing Technology (ACTs), and direct technology licensing through its technology transfer offices. However, a heretofore underutilized tool is the Other Transaction Authority (OTA).

An Other Transaction (OT) is a special mechanism used by federal agencies for obtaining or advancing R\&D or prototypes; it is not a contract, grant, or cooperative agreement, and there is no statutory or regulatory definition of “Other Transaction.” OTA is valuable in cases where the government needs to obtain R\&D and prototypes from commercial sources, but the companies equipped to provide them are unwilling or unable to comply with the government’s procurement regulations. The government’s procurement regulations and certain procurement statutes do not apply to OTs; thus, OTA gives agencies the requisite flexibility to develop agreements tailored to a particular engagement. 

While the Energy Policy Act of 2005 granted OTA to DOE at the agency level, it failed to authorize the labs to use it.  OTA could be offered as a unique authority provided by DOE to the labs that can enable HEP to showcase a more effective model for technology transition— but drive it at the local laboratory level, where the interaction with industry is vital for success. 
OTA is an ideal mechanism to help labs better identify market needs and become more valuable to the private sector because it can establish a formalized relationship where both parties have “skin in the game” early on in the research process, so markets can be better understood for deployment of technology.  For example, under Stevenson Wydler, laboratory partners cannot be funded by means of a CRADA. Thus, CRADA partners, if they are to be funded, must enter procurement subcontracts with the laboratories, which not only introduces another complex agreement to the partnership, but potentially disparate terms and conditions. Because OTA is merely defined in the negative, an OT can be crafted to function much like a CRADA, but without the prohibition of funds-out.  

Additionally, OTA has been used successfully in other agencies such as the Department of Defense to construct consortium arrangements with multiple parties.  These consortia establish an interested and engaged set of stakeholders early on in a technology lifecycle, which can then lead to a set of potential partners for further collaboration or even vendors who could manufacture the technology in the future.  

Current mechanisms (e.g., CRADAs and SPPs) don’t engage the deployment / commercialization community early enough in the development process. As a result, industry does not have enough incentive to be part of the process in defining the deployment model.  Having agency funding tied to OTA and a funds-out collaborative agreement keeps the critical researchers involved, as well as a partner— but can call out specific performance deliverables related to transition that the agency funding the OTA could request.  Furthermore, the OTA allows for negotiation of onerous federal contract provisions that often keep companies from doing business with the federal government.  Therefore, HEP could be “open for business” with non-traditional federal contractors who otherwise would not be willing to do business with DOE.  

Additionally, a collaborative mechanism with bilateral flow of funds better ensures the risk of failing to commercialize the technology is not carried exclusively by the industry partner, but also by the laboratory.  HEP could fund projects that authorize use of funds-out collaborative agreements with industry partners that then leverage federal funds as well as industry in-kind contributions.  The OTA includes provisions and milestones for performance that can include: prototype development, delivery, market scope definition, deployment models, etc., which are elements not included in R\&D subcontracts. With requirements as part of the funding mechanism (along with provisions with respect to intellectual property), both parties are focused early in the research process on moving a national lab technology to deployment.

Because OTA affords flexibility not available in other agreements, HEP can set expectations about technology transition but drive the activity at the research/partnership level rather than the agency level.  This will enable tailoring of an arrangement most suitable for the technology and the market in which the technology will be deployed, with terms and conditions that reflect the risks and potential reward chain available to the industry partner. 

\subsection{Establishing an Entrepreneurial Leave Program}
An Entrepreneurial Leave Program (ELP) allows employees to take a leave of absence or separation from the laboratory in order to start or join a new company. ELPs encourage startup activities by reducing the risks faced by the employee entrepreneur. Some elements of an ELP may include business preparation / training, a means for licensing laboratory IP, continuity of health benefits during leave, and a mechanism for returning to work.

ELPs are not implemented consistently across the DOE complex; some laboratories have ELPs while others do not. A harmonized ELP could help to attract and retain talented staff to the national laboratories by drawing entrepreneurial candidates to the laboratories and providing a path for returning to work. By facilitating a return to the laboratory, DOE brings these unique entrepreneurial experiences back into the Federal R\&D network. A survey of lab ELP policies is available~\cite{atherton_2015}.

%%%%%%%%%%%%%%%%%%%%%%%%%%%%%%%%%%%%%%%%%%%%%%%%%%%%%%%%%%%%%%%%%%%%%%%%%
% example figure

%\begin{figure}
%\begin{center}
%\includegraphics[width=0.40\hsize]{xxx}
%\end{center}
%\caption{xxx}
%\label{fig:xxx}
%\end{figure}

%%%%%%%%%%%%%%%%%%%%%%%%%%%%%%%%%%%%%%%%%%%%%%%%%%%%%%%%%%%%%%%%%%%%%%%%%%%

%%%%%%%%%%%%%%%%%%%%%%%%%%%%%%%%%%%%%%%%%%

%  If you would like to use BibTEX for the bibliography, please feel free to do so.  It is not required.

%  To use BibTeX,

%    1.  uncomment the following two lines, 
%    2.  comment out everything below from  \begin{thebibliography}{99}   to \end{thebibliography).
%    3.  create the file  myreferences.bib, and process this file in the usual way

\bibliographystyle{JHEP}
%\bibliography{refs}
%\bibliography{myreferences}  % file myreferences.bib

%%%%%%%%%%%%%%%%%%%%%%%%%%%%%%%%%%%%%%%%%
%example bibliography

\appendix
\section{Appendix A: Fermilab Commercialization Process of HEP Technologies by Spin-offs or Start-ups}
\input{sections/Fermilab_process}

\section{Appendix B: KEK Commercialization Process of HEP Technologies by Start-ups}
\input{sections/KEK_process}

\end{document}

%% file: workshopsymbols.tex
\def\beq{\begin{equation}}
\def\eeq#1{\label{#1}\end{equation}}
\def\eeqn{\end{equation}}

\newenvironment{Eqnarray}%
   {\arraycolsep 0.14em\begin{eqnarray}}{\end{eqnarray}}
\def\beqa{\begin{Eqnarray}}
\def\eeqa#1{\label{#1}\end{Eqnarray}}
\def\eeqan{\end{Eqnarray}}

\let\bar=\overbar

\def\lsim{\mathrel{\raise.3ex\hbox{$<$\kern-.75em\lower1ex\hbox{$\sim$}}}}
\def\gsim{\mathrel{\raise.3ex\hbox{$>$\kern-.75em\lower1ex\hbox{$\sim$}}}}

\def\del{\partial}
\def\Dslash{\not{\hbox{\kern-4pt $D$}}}
\def\dslash{\not{\hbox{\kern-2pt $\del$}}}
\def\pslash{\not{\hbox{\kern-2pt $p$}}}
\def\ETmiss{\not{\hbox{\kern-4pt $E$}}_T}

\def\Dlr{\mathrel{\raise1.5ex\hbox{$\leftrightarrow$\kern-1em\lower1.5ex\hbox{$D$}}}}

\def\MSB{{\bar{M \kern -2pt S}}}
\def\msb{{\bar{\scriptsize M \kern -1pt S}}}

\def\drb{{\bar{\scriptsize D \kern -1pt R}}}

%% file: sections/Fermilab_process.tex
\subsection{IP Royalty}
The minimum royalty percentage required by US statute to be given to an inventor is 15\%.  Technology transfer legislation also mandates that each federal agency that spends more than \$50 million per fiscal year on R\&D shall develop a cash awards program to reward researchers for their innovations.  As a Government Owned Contractor Operated (GOCO) facility, Fermilab has greater freedom in the way it distributes royalties; however, its current policies remain close to the federally mandated regulations. FRA inventor employees share 50\% of net royalties (after direct expenses) on annual net royalties below \$10,000. If annual net royalties are \$10,000 or greater, the inventor employees share \$5,000 of the first \$10,000 in net royalties and 25\% of net royalties in excess of \$10,000. In cases for which Fermilab negotiates recovery of patent costs from the licensee as a distinct payment (e.g. exclusive licenses), patent expenses are not deducted from gross royalties prior to sharing of the proceeds.

\subsection{Technology License}	
A technology license is a contract that details rights and obligations. Licenses can differ greatly depending on the type of technology and the type of rights requested. For more information, please see:

https://partnerships.fnal.gov/licensing/licensing-guidelines/

https://www.autm.net/AUTMMain/media/ThirdEditionPDFs/V4/\\TTP\_V4\_AnatomyLicense.pdf

There is a term “U.S. Government retained license” in the first link above. Through a series of landmark legislative acts in the 1980s (e.g. the Bayh-Dole Act), universities and contractors operating national laboratories were authorized to retain title to inventions created using federal funding, pursue patent protection, and license those inventions to third parties. These regulations also stipulate government rights, including the right of the US Government to use the inventions for its own internal purposes (U.S. Government retained license) and Licensee obligations, for example to substantially manufacture in the U.S. for U.S. products for exclusive licenses. For an interesting article on this subject please check: https://www.bnl.gov/techtransfer/docs/TTWG-LicensingGuide.pdf

\subsection{Process of Requesting a License}	
FRA Management will ensure widespread notice of the availability of technologies suited for transfer and opportunities for exclusive licensing through the Fermilab website and will invite comments submitted by interested parties. In addition, notice of exclusive licenses may be provided by other media such as: publication, participation in conferences, exhibitions, or competitive solicitations.

If Fermilab’s Management determines that there has not been a sufficient period of notice given to establish fairness, or that based upon the evidence and argument submitted in writing by a third party that it would not be in the best interest of the general public, FRA/Fermilab and the United States Department of Energy to grant the exclusive or partially exclusive licenses, a written notice of such decision shall be documented and provided to the third party.

FRA Management reserves the right to extend the notice period if it determines the circumstances warrant additional opportunities for notice and comment are necessary to meet its obligation of fairness of opportunity under its DOE Contract.

An inventor’s team can portray credibility of commercial intent by having a commercialization plan. The inventor understands that it usually takes a team with diverse skills (technical, financial, legal, etc.) to commercialize an invention. It also requires time and resources. The commercialization plan will outline the resources needed, how these resources will be obtained and the timeline of how they will be deployed to reach well defined milestones.

The Laboratory requires a commercialization plan from all its prospective licensees before licensing its intellectual property. The plan better enables the Laboratory and the prospective licensee to determine and characterize the business fit between their interests. The commercialization plan also provides information for the Laboratory to use in assessing the prospective licensee’s intent and plan to achieve commercial use of the Laboratory Intellectual Property. Finally, the plan provides the Laboratory input on its calculation of value of the intellectual property to both Parties and the type of rights that the startup needs in the license (e.g. what type of Field of Use? Or what type of Territory?)~\cite{sauers1_2016}.

Fermilab licensing royalties are comparable to those charged by universities, other research organizations and the private sector. Licenses usually require an upfront, nonrefundable payment, royalty payments based on sales, and a minimum annual royalty. The fees will vary depending on the number of patents/copyrights licensed, the demand for the technology and the exclusivity of the license. Licensees obtaining foreign rights may be asked to pay the cost of preparing, filing and prosecuting foreign patent applications, and the maintenance of all resulting foreign patents~\cite{}.

FRA can accept equity in lieu of up-front payments in the license. While FRA can accept equity as consideration for grant of IP rights, it favors phantom stock provisions over equity. Phantom stock may generally take the form of a milestone payment tied to company valuation. Phantom stock may be useful when working with startup companies without the ability to commit significant cash outlays for both developmental and licensing costs (e.g. patenting costs and/or licensing fee).

\subsection{OPTT Process to Secure Authorization}
%What is the process that OPTT needs to follow to secure authorization in order to provide a license to one of its inventor employees?
%"The laboratory must obtain the approval of the contracting officer prior to any assignment, exclusive licensing, or option for exclusive licensing, of Intellectual Property to any individual who has been a Laboratory employee within the previous two years or to the company in which the individual is a principal" Contract No. DE-AC02-07CH11359 CLAUSE I.105 - DEAR 970.5227-3 TECHNOLOGY TRANSFER MISSION DEVIATION: AL-2006-10 part d, Conflicts of Interest
Many startups would require an exclusive license (at least in a relevant field of use) in order to secure funding from investors. If that is the case and an exclusive license is requested, FRA will need to make sure that there is “fairness of opportunity” for other entities to be able to compete for the license. Fermilab’s OPTT will ensure widespread notice of the availability of the technology suited for transfer through the Fermilab website. In addition, notice of exclusive licenses may be provided by other media such as: publication, participation in conferences, exhibitions, or competitive solicitations~\cite{sauers2_2016}.

FRA may grant exclusive or partially exclusive licenses in any invention only if the invention has been published for licensing for a period of at least thirty (30) days. After the thirty (30) day notice period, FRA Management will determine whether:
\begin{itemize}
    \item The interests of the general public, FRA/Fermilab and the United States Department of Energy will be best served by the proposed license; and,
    \item It does not appear that the desired practical or commercial application has been or will be achieved on a nonexclusive basis; and that,
     \item Exclusive or partially exclusive licensing is a reasonable and necessary incentive to call forth the risk capital and expenses necessary to bring the invention to the point of practical or commercial application.
\end{itemize}
Besides making sure that there is fairness of opportunity, FRA will also need to make sure that there is no conflict of interest, as described below.

Conflict of interests need to be reviewed prior to providing a license to an employee inventor. Examples of conflict of interest may arise when there is an invention related to a collaboration, was created under contract, or is related with the employee’s prior employment.  

DOE’s Fermi Site Office Contracting Officer needs to sign-off any requests for licenses by FRA employees and any individual who has been a Laboratory employee within the previous two years or to the company in which the individual is a principal.

The Laboratory must ensure it both has the rights it intends to license and is able to provide those rights to the requester. As a result, the Laboratory may take steps to perfect its rights prior to a proposed licensing arrangement. In addition, intellectual property may be encumbered by existing license agreements or options to license. 

\subsection{Educational Programs for FRA Inventors}
FRA is a partnership between the University of Chicago and the Universities Research Association (URA). The fact that University of Chicago is part of the management of FRA gives FRA employees the ability to take advantage of many educational and entrepreneurial resources provided by University of Chicago. For example, FRA entrepreneurs can participate in:
\begin{itemize}
\item Polsky I-CORPS (https://polsky.uchicago.edu/programs-events/polsky-i-corps/),
\item Collaboratorium (https://polsky.uchicago.edu/programs-events/collaboratorium/),  
\item free non-credit classes like the Intro to Venture Capital \\ (https://polsky.uchicago.edu/event/application-deadline-intro-to-venture-capital/).
\end{itemize}

FRA employees may also participate in:
\begin{itemize}
\item DOE’s Energy I-CORPS (https://energyicorps.energy.gov), 
\item FRA’s Practicum (https://news.fnal.gov/2020/03/fy20-entrepreneurship-\\and-commercialization-practicum-new-schedule/), 
\item and/or discuss with OPTT about other possible options.  
\end{itemize}

\subsection{Accelerator Programs for FRA Inventors}
Once you secure a license to the technology, you’re free to join any startup accelerator in the nation that you want. 
DOE hosts three accelerators that specialize in energy and hard tech: 
\begin{itemize}
\item Chain Reaction Innovation at Argonne National Lab \\
(https://chainreaction.anl.gov/program/) 
\item Innovation Crossroads at Oak Ridge National Lab \\ (https://innovationcrossroads.ornl.gov)
\item Cyclotron Road at Berkeley Lab (https://cyclotronroad.lbl.gov)
\end{itemize}
All FRA entrepreneurs are encouraged to learn about these programs as part of their commercialization journey.

For FRA inventors who wish to take time-off to work in their startup,
FRA does not currently have a standard Entrepreneurial Leave Program but has and will continue to entertain requests for sabbatical or reduced employment hours on a case-by-case basis.

%% file: sections/KEK_process.tex
\subsection{IP Royalty}
The inventors get a percentage (after patent expenses) of the revenue received by KEK. KEK inventors share 40\% of net royalties personally or they can use 65\% to 95\% of net royalties as research fund. They can also select both the personal and the research options at the same time. In addition, they will receive around 15,000 Japanese Yen (the exact amount depends on number of claims and inventors) when the patent is admitted at the Patent office.

\subsection{Process of Requesting a License}
If an inventor starts a venture company which utilize the invention, KEK will admit and give several supports to the company, such as office space, exclusive patent license, and permission of using KEK name etc. 
Many inventions are shared patents with collaborating companies. In this case, KEK request non-implementation compensation to the company, or ask all the patent-related fee to the company or negotiate paid transfer of the patent.

As for the independent patent, IPPO and inventors try to find interested companies. Although it is not easy to find interested company, companies working around the inventor may show interest in some case.

KEK can accept equity as consideration for grant of IP rights but there is no such case happened up to now.

If a venture company is established by KEK employees for the technology transfer of their invention, KEK can support the company after examination. The support term is 5 years and it can be extended another 5 years if it is admitted as necessary. The supported venture company can use the name of 'KEK venture' and KEK space/equipment. They can get exclusive license of their invention and preferential treatment.

Examples of recent IPs licensed to companies or jointly invented are shown below:
\begin{itemize}
\item IPs related to Non-Evaporable Getter (NEG) for vacuum pump, 
\item IPs related to super-conducting magnet,
\item IPs related to super-conducting RF accelerating chamber,
\item IPs related to ultra-low temperature device,
\item IPs related to making cathode/target material,
\item IPs related to making SOI pixel detector,
\item Low power oscillator circuit.
\end{itemize}